# The History of the Square Kilometre Array (SKA) Born Global

Ron Ekers[1]
*Affiliation, CSIRO*
*Address, Australia*
*E-mail:* ron.ekers@csiro.au

A brief review of the history of the Square Kilometre Array (SKA) from its pre 1990 roots and the global vision which emerged, at the VLA 10th anniversary meeting in 1990, to the major international project we have today. I comment on the evolution of the science and the technology that has occurred during this period. Finally, we can ask: "What have we learned?"

[1] Speaker

## 1. Introduction

Richard Schilizzi became the first SKA project director in 2002 so it seemed very appropriate at Richard's retirement to review the early history of the SKA.

I have tried to trace the historical developments as objectively as possible but I have also added my personal views of these developments and commented on lessons learned.

## 2. Pre 1990

Barney Oliver and John Billingham [1] made a detailed design of a 16km diameter array of 1000 100m dishes working at cm wavelengths which would be a realistic effort aimed at detecting extraterrestrial intelligent life. Called project Cyclops this was not dissimilar to the current SKA concept but about an order of magnitude beyond SKA in cost ($10 billion) and sensitivity.

In the period 1986-9 the Canadians proposed a Radio Schmidt telescope with emphasis on surveys through the wide field of view. Peter Dewdney's 100 x 12m antennas were discussed at the 1989 Penticton Radio Schmidt Workshop. This concept extended the synthesis imaging paradigm by using large numbers of small dishes to create an all sky survey telescope playing a role analogous to the 48" Palomar Schmidt optical telescope and its all sky survey. However, the vision did not include sufficient sensitivity to do new science and the proposal languished.

Between 1988 and 1990 Robert Braun, Ger de Brujn and Jan Noordam proposed a Dutch extragalactic HI telescope. Their key idea was to build a telescope with enough collecting area to detect HI at high redshift. Independently, but at about the same time, Govind Swarup in India was already planning a next generation cm radio telescope based on the Giant Metre Radio Telescope (GMRT) [2] which was already under construction. In 1991 Swarup published his proposal for an International Radio Astronomy telescope (ITRA) which included 160 75m dishes, centrally concentrated with baselines up to 200km [3].

### 2.1 Exponential growth in science

At the 1990 URSI General Assembly in Prague, in the Commission J tutorial Ekers argued for the continued exponential growth of radio telescopes based on the discovery arguments of Derek de Solla Price [4].

Harwit [5] showed that most important discoveries in astronomy result from technical innovation. It had already been well established that most scientific advances follow technical innovation in other areas of science [6]. De Solla Price applied quantitative measurement to the progress of science (scientometrics) and reached the conclusion that most scientific advances follow laboratory experiments. His analysis also showed that the normal mode of growth of science is exponential [4]. A rather simplified conclusion to draw from this is that any field which has not maintained an exponential growth has now died out, so current active research areas are all still in an exponential growth phase. Furthermore, to maintain the exponential the continual introduction of new technology is required since just refining existing technology plateaus out.

Figure 1 plots the point source continuum sensitivity of telescopes used for radio astronomy since the first discovery of extra-terrestrial radio emission in 1940. It has been exponential with an increase in sensitivity of $10^5$ since 1940, doubling every three years. Also in this case we can see particular radio telescope technologies reaching ceilings and new technologies being introduced e.g., the transition from huge single dishes to arrays of smaller dishes in the 1980s.

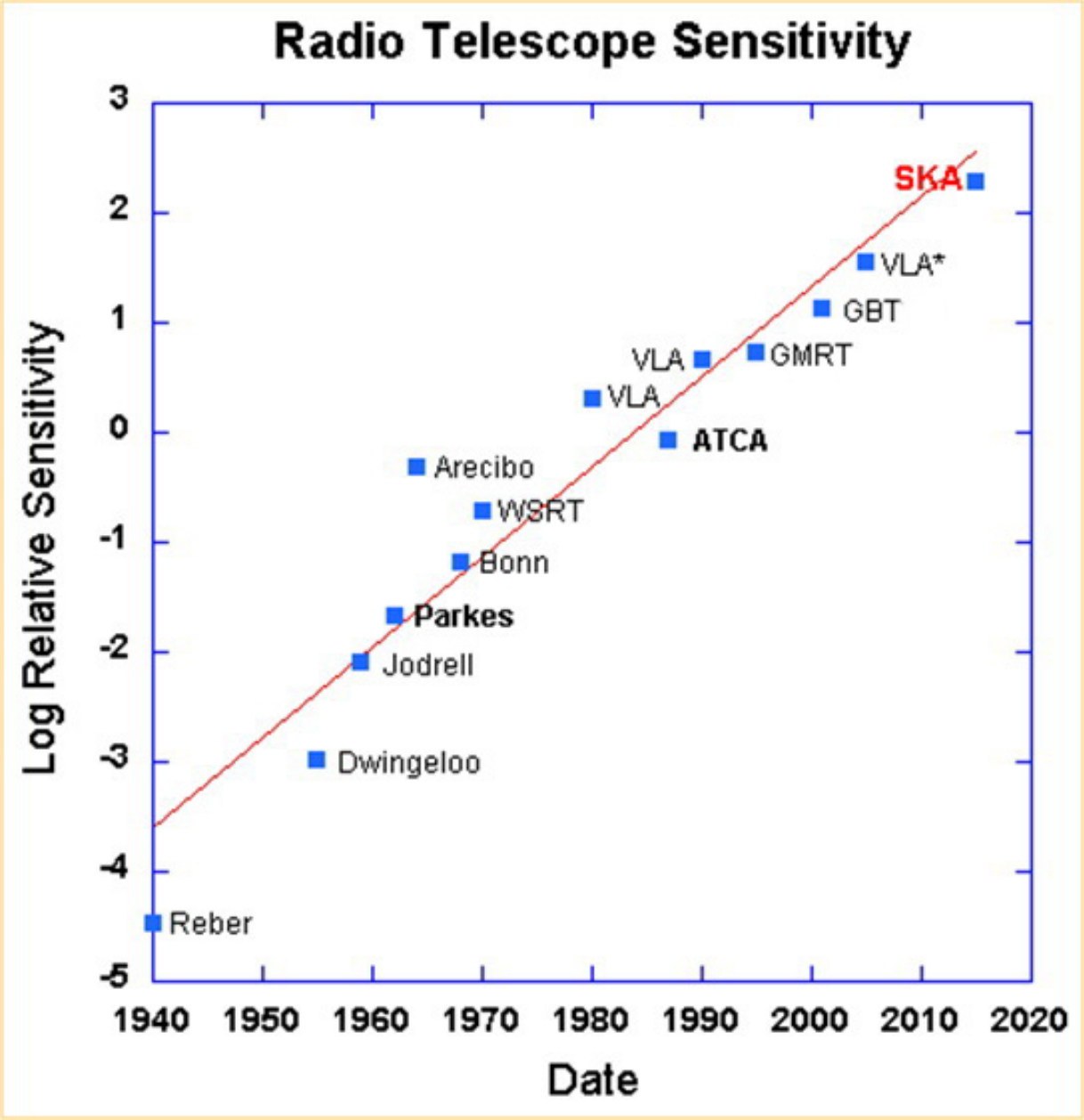

**Figure 1**: Radio Telescope Sensitivity vs. time. Points are the relative continuum point source sensitivity when the telescopes were built, or after major upgrades. VLA* is the EVLA upgrade, now named the Jansky VLA.

## 3. The VLA 10th Anniversary Meeting

IAU colloquium 131, *Radio-Interferometry*, was held in Soccorro, New Mexico, USA, 8 October 1990, to celebrate the first 10 years of observations with the VLA. At the meeting, Jan Noordam from the Netherlands talked to Peter Wilkinson from Jodrell Bank, UK, about the Netherlands Foundation for Radio Astronomy (NFRA) large HI telescope proposal. In his talk at the meeting Wilkinson [6] included the science case for a radio telescope with $1km^2$ of collecting area to study HI in galaxies at high redshift. He called this telescope the Hydrogen Array.

In the same session Yurij Parijskij from the Pulkovo Observatory, Soviet Union, described the historical evidence for exponential improvements over time of many factors which contribute to radio telescope performance: collecting area, resolution, sensitivity. Parijsky [8] argued for a telescope with 1 $km^2$ total area by the year 2000 to beat the exponentially increasing RFI threat.

A more colourful version of this early history is given by Jan Noordam in these proceedings.

## 4. Achieving the Vision International Collaboration

### 4.1 Born Global

From its beginning the SKA was conceived as an international project. The scientists and engineers involved were accustomed to working together and they shared a radio astronomy culture with an open Sky policy and shared technology development. At an early stage links were made to existing International organizations including: IAU, URSI, OECD, and the EU.

Through such international collaboration we can build facilities which no single nation can afford. By coordinating efforts we can avoid wasteful competition and the broader knowledge base and cross fertilisation will stimulate innovation. The final decision to fund such large projects will be taken because of its value for wealth creation in the collaborating nations. Science can provide the shared vision needed to pursue such collaborations.

### 4.2 Open Sky Policy

Almost all radio observatories operate with an "open sky" model in which access is not limited to scientists from country or organization that operates the telescope. This is usually justified on the basis that it guarantees the best science return with the facility whereas guaranteed access rights for scientists from funding nations favours the individual scientists more than the funding nations. The facilities still get the recognition regardless of the nationality of the user and with an open skies policy it is easier to set up the large teams conducting surveys which can be made available to the entire community.

## 5. Brief History of the SKA

In the next part of this presentation I will expand on some of the items in the following timeline:

1988 Independent suggestions for a Large Radio Telescope
1990 10th anniversary of VLA – the visions merge
1993 URSI GA Kyoto resolution to establish the Large Telescope Working Group
1994 IAU forms its Future Large Scale Facilities Working Group
1996 OECD Global Science Forum included radio astronomy activities
1997 MoA for technology collaboration
1998 “SKA” name adopted
2000 International MoU signed
2001 Logo competition
2003 International Project Office established
2004 Rawlings and Carilli published the new science case
2006 Site down-selected to Australia and South Africa
2008 SKA Project Development Office (SPDO) formed
2009 Agencies SKA Group formed
2011 SKA becomes a legal entity.

### 5.1 URSI General Assembly, Kyoto, September 1993

In September 1993 the International Union of Radio Science (URSI) established the Large Telescope Working Group to begin a worldwide effort to develop the scientific goals and technical specifications for a next generation radio observatory.

At the URSI General Assembly in Kyoto in September 1993, Commission J resolved to form a Large Telescope Working Group to consider:

*a) The strong scientific case for a new, internationally accessible radio telescope with one or two orders of magnitude greater sensitivity than that of any existing or planned facility;*

*b) The need for innovative technical developments to realize such a facility at an affordable price;*

*c) The likely need for international collaboration to allow realization of this facility.*

And resolved to appoint a working group with the following terms of reference:

*1) To explore the range of scientific problems to be addressed by the instrument.*

*2) To discuss the technical specifications and general design considerations needed to maximize the scientific return of such a facility.*

*3) To identify and, in so far as possible, resolve the major technical challenges to realisation of an affordable radio telescope with the required sensitivity.*

Membership: Robert Braun (NFRA, Netherlands), Ron Ekers (ATNF, Australia), Lloyd Higgs (DRAO, Canada), Yuri Parijski (SAO, Russia), Wolfgang Reich (MPIfR, Germany), Wu Shengyin (Beijing Obs, China), Govind Swarup (TIFR, India), Dick Thompson (NRAO, USA), Peter Wilkinson (NFRA, UK). Subsequent meetings of the working group provided a forum for discussing the technical research required and for mobilising a broad scientific community to cooperate in achieving this common goal [9].

### 5.2 IAU Future Large Scale Facilities Working Group, 1994

Following a joint discussion on Future Large Scale Facilities in Astronomy held at the XXIInd IAU General Assembly in the Hague, August 1994, a working group of the IAU executive was established to explore international implications of the development of large astronomical facilities at all wavelengths. This broader concept had been triggered by the interest and success of the radio telescope initiative (SKA). This IAU working group is still active. Its tasks are:

*(i) To promote the early discussion and dissemination of information on potential large scale astronomical projects;*

*(ii) to make and maintain an inventory of planned large scale projects in astronomy;*

*(iii) to further contacts and cooperation between different projects;*

*(iv) to identify areas of duplication and areas where clearly desirable efforts are lacking.*

### 5.3 OECD Global Science Forum, 1996

In 1996 the OECD Mega Science Forum, later called the Global Science Forum, looked at international big science projects in astronomy. In 1998 the OECD Mega Science Forum set up a Task Force on Radio Astronomy which included astronomers, regulators, and the satellite communications industry. The terms of reference were to investigate methods to provide international protection from satellite communications for future large radio telescope projects built at a few designated locations [10].

In 2003 a Global Science Forum on Astronomy looked at the opportunities for global collaboration on planning processes for future facilities. While many important issues were raised, the attempt at international coordination of the various independent national programmes failed.

### 5.4 Cooperative agreement, 1997

In 1997 in an initiative inspired by Harvey Butcher (NFRA) eight institutions from six countries (Australia, Canada, China, India, the Netherlands, and the U.S.A.) signed a Memorandum of Agreement to cooperate in a technology study program leading to a future very large radio telescope.

### 5.5 SKA gets a name, 1998

At the July, 1998, large radio telescope science meeting in Calgary it was agreed to adopt a single name for the project. The various prototypes under planning or construction would keep their own distinct names determined by local priorities, but it was agreed that everyone, worldwide, would use the same name in referring to the international project.

A vote was held by those present at the meeting, and in a follow-up electronic discussion. A majority chose the name *Square Kilometre Array, SKA*, over the many others that had been suggested, including; OSKAR, 1kT, SKAI, ITRA, KARST, SLA, VLRT, SKIRT, Argo, NGRO, NGAT,

### 5.6 International MOU signed, 2000

On August 10, 2000, at the IAU General Assembly in Manchester, UK, a Memorandum of Understanding to establish the International Square Kilometre Array Steering Committee (ISSC) was signed by representatives of eleven countries. The ISSC had 18 members representing 11 countries: Europe (UK, Germany, Netherlands, Sweden, Italy, Poland) (6), United States (6), Canada (2), Australia (2), China (1), India (1) and 2 at large members.

This was eventually superseded by a Memorandum of Agreement to Collaborate in the Development of the Square Kilometre Array which came into force on 1 January 2005 and was extended until 31 December 2007. It made provision for the expansion of the Steering Committee to 21 members (7 each for Europe, USA, and the Rest of the World) and the establishment of the International SKA Project Office.

### 5.7 Logo competition, 2001

An international competition was conducted to choose a logo for the SKA. The adopted solution (Figure 2 centre) merged two design concepts to capture both the radio wave and the global vision.

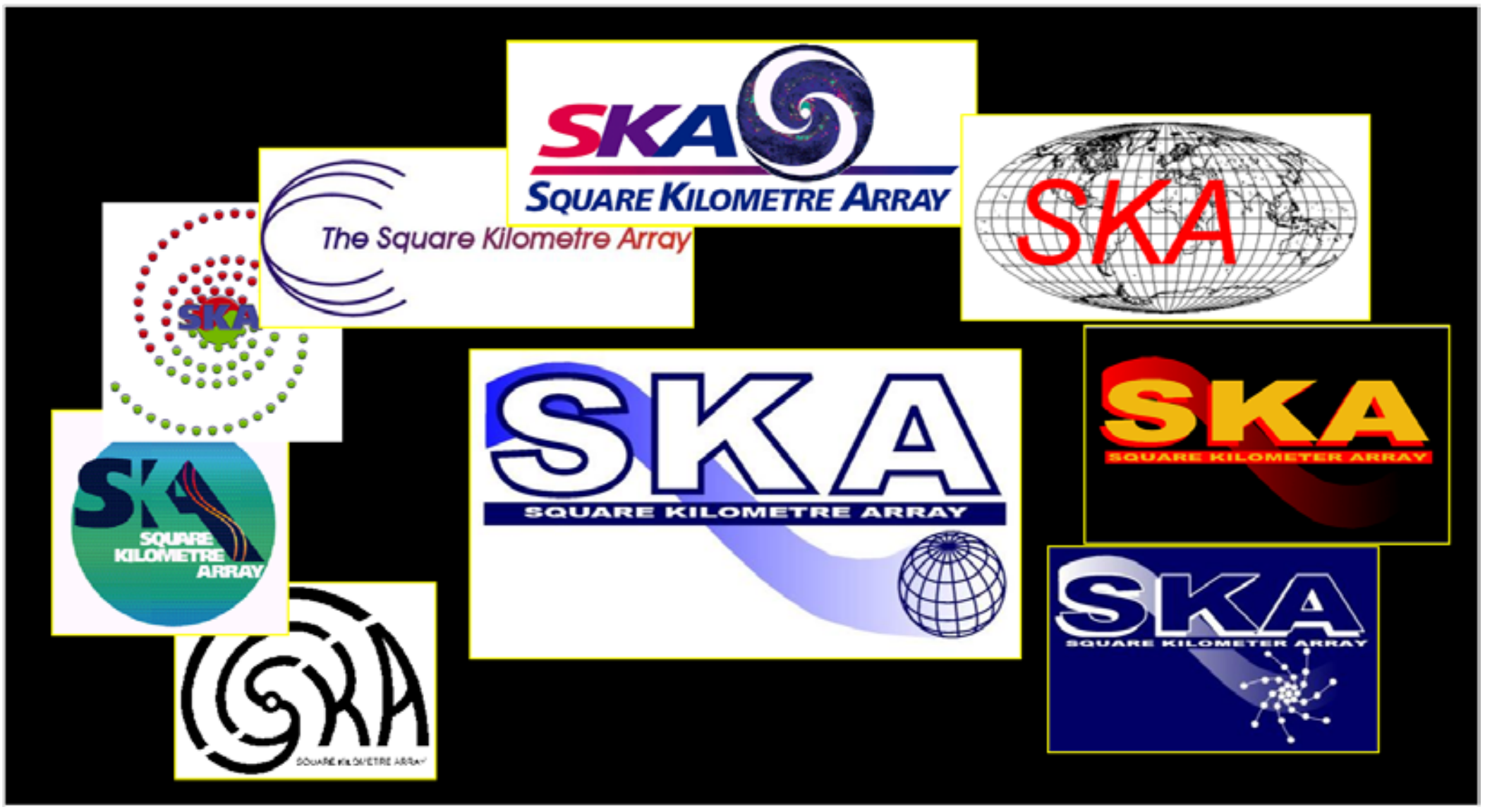

**Figure 2:** SKA logos entered into the International competition

### 5.8 SKA International Project Office, 2003

The office was established in 2003; Richard Schilizzi was appointed as the first SKA International Project Director and was initially based at NFRA in the Netherlands. In 2008 the Project Office moved to Manchester.

## 6. The Technology Challenge

The SKA has always been aperture arrays at lower frequencies and dish arrays at higher frequencies with the transition frequency to be determined by technology development. The technology challenge has always been to provide a square kilometre of aperture at an acceptable cost. To do this it was envisaged that the SKA would have to make a revolutionary break with current radio telescope design. New technologies involving large scale integration of transistor amplifiers into complex systems which can be duplicated inexpensively, and the rapidly increasing capability to apply digital signal processing at high bandwidth now make it possible to obtain this aperture at an affordable cost. Institutions participating in the SKA are now designing and building prototype systems and the key technologies will be determined from these.

### 6.1 Technology evolution

The development of many innovative technologies has been stimulated by the extreme requirements of the SKA and by the cross fertilisation between many international groups. Exploring these developments in detail is beyond the scope of this paper but I have tried to summarise the highlights, and indicate the dates and locations of the international SKA meetings where some of these concepts first emerged. Many individuals were involved and in some cases ideas arose independently from multiple sources. It is my intention to better document these developments so any feedback on these highlights would be welcome.

1990 (Socorro)
– arrays of large dishes
1994 (IAU den Hague)
– Dense Aperture Array (AA) tiles – observing all sky with no moving parts [11]
– Canadian very long focal length telescope (LAR)
– Chinese deformable spherical telescope (FAST)
1996 (Delft)
– Hierarchical arrays and hierarchical beam forming
– Multiple simultaneous beams
– Phased Array Feeds (PAFs)
1997 (Sydney)
– RFI mitigation – adaptive nulling
1997 (Oort workshop Leiden)
– The Low Frequency Array (LOFAR)
1997-1999 (SETI workshops, California [11])
– Large N small D arrays for cost minimization
– Very wideband feeds and receivers
– Alan Telescope Array (ATA)
1999 (Dwingeloo)
– Luneburg lens
– Large N small D arrays for image quality [13]
2000 (Manchester)
– Deep Space Network Array
2003 Geraldton
– Dishes with PAFs to move cost into components which will get cheaper
2004 (Capetown)
– Cylinders
2005 (Heathrow)
– Beginning of the Reference Design

## 7.The Science

### 7.1 Science Case

The first detailed analysis of the science cases for the SKA was written by Russ Taylor & Robert Braun in 1999 [13]. A revised and updated science case was commissioned at the SKA steering committee meeting in Bologna in 2002. Chris Carilli and Steve Rawlings edited contributions which were published in 2004 [15]. These well documented science cases have strongly influenced the design of the SKA and have generated interest in a much broader community.

### 7.2 Evolution of the Science case

The need to measure HI at cosmological distances is unchanged from the beginning of the SKA concept of a Hydrogen Array but what we use the HI for has changed a lot! Initially it was proposed to use it to detect Zeldovich pancakes but now now it is proposed to use HI as a probe large scale structure. As quoted by Peter Wilkinson [7] *The encyclopaedia of the Universe is written in very faint (21-cm) typescript, to read it one requires a very large lens.*

Pulsars were included in the earliest proposals since sensitivity is always at a premium and pulsar research has slowly gained in importance over time. There are now many more proposed General Relativity tests, partly stimulated by the discovery of the double pulsar [16]. The use of pulsar timing for gravitational wave detection has been further stimulated by strong interest from the physics community and by the complementary proposals such as LIGO and LISA. However pulsars need high, instantaneous Ae/T which can't always be traded for survey speed and this has an important impact on the design.

Detection of the epoch of re-ionisation was first mentioned as an SKA goal in 2002 based on predictions made in 1997 [19]. By 2004 the plausibility and implications of detection were being discussed in many papers, e.g. [20].

The recent evolution of the science case to emphasise a few key science goals needs to be kept in perspective. It is essential to have a short list of science goals in order to promote the SKA and it is valuable to have a small number of test cases which can be used for detailed design considerations. However, it is the exploration of the unknown that is likely to make the most important new discoveries and will raise new questions and new problems [17].

## 8.What have we learned?

Visionary projects take a long time! It is more than 22 years since the first clear SKA vision emerged.

### 8.1 Collaboration

The radio astronomy culture has played a critical role in the development of this project. This is a combination of the open policies for access to facilities and the sharing of technology. Also the strong link between engineering and science, which is normal in this field, allows interplay between technical innovation and science opportunities.

Initially we had many questions about the lack of a suitable international structure. There was no organization like ESO or CERN for radio astronomy. The scientific unions were able to facilitate communication but they don't build telescopes. New international structures have now been created with strong support from Governments that recognise the value of the international collaboration as well as the science.

Broad community involvement in the SKA has been achieved but this has also resulted in a failure to control the scope of the project and a dependence on decision making by consensus. Galileo (1612): *In questions of science the authority of a thousand is not worth the humble reasoning of a single individual* [22] [21].

### 8.2 Expectations vary with stakeholders

The astronomers want the best science and will be motivated by the scientific observations which new instruments make possible. While it is essential to have the science drive the technology and provide the technical challenge, we should be wary because the key science evolves on a faster time scale than a major facility can be built.

Governments want wealth creation, international links and prestige. As outlined in the COST strategic workshop 2010 [18], there are benefits of research infrastructure beyond science that can result from the SKA and other large scale infrastructure research projects. In the case of the SKA this is expected in at least in four areas: Information and communication technology, renewable energy, global science-industry-government linkages and human capital development.

Industry wants contracts and they need funded opportunities to do the cutting edge R & D that will generate future commercial opportunities.

### 8.3 The Pathfinders and the SPDO

The scale of the pathfinders has been an issue. Astronomers want useful steps hence big pathfinder projects. Technology development needs small, diverse R&D activities but without the astronomical drivers there is no guarantee that the appropriate technology is developed. Now some of the pathfinders are major facilities in their own right we see issues where national interests may not be completely aligned with the international vision. The early attempt at centralisation to unify the project has lead to paper designs and has alienated some of the key engineers who want to work on real hardware.

Overall, the SKA triggered pathfinders, LOFAR, MWA, MeerKat, ASKAP, FAST, have been a huge success and have revitalised radio astronomy. A new age of radio astronomy has been born but we still have a challenge to realise the full SKA vision.